\documentclass[iop,revtex4]{emulateapj} 

\usepackage{graphicx,subfigure}
\usepackage{float}
\usepackage{booktabs}
\usepackage{color}
\usepackage{amsmath}
\usepackage{enumerate}

\slugcomment{Submitted to the Proceedings of the Astronomical Society of the Pacific}

\newcommand{\comm}[1]{}

\shorttitle{Enhanced spectroscopy with extreme adaptive optics}
\shortauthors{N. Jovanovic \it{et al.}}

\begin{document}
\title{Enhancing stellar spectroscopy with extreme adaptive optics and photonics} 

\author{N. Jovanovic\altaffilmark{1,2}, C. Schwab\altaffilmark{2,3}, N. Cvetojevic\altaffilmark{3,4,5,6}, O. Guyon\altaffilmark{1,8,9} \& F. Martinache\altaffilmark{10}}
\altaffiltext{1}{National Astronomical Observatory of Japan, Subaru Telescope, 650 North A'Ohoku Place, Hilo, HI, 96720, U.S.A.}
\altaffiltext{2}{Department of Physics and Astronomy, Macquarie University, NSW 2109, Australia}
\altaffiltext{3}{The Australian Astronomical Observatory  (AAO), Level 1, 105 Delhi Rd, North Ryde, NSW, 2113, Australia} 
\altaffiltext{4}{Centre for Ultrahigh bandwidth Devices for Optical Systems (CUDOS)}
\altaffiltext{5}{Institute of Photonics and Optical Science (IPOS), School of Physics, University of Sydney, Sydney, NSW, Australia}
\altaffiltext{6}{Sydney Institute for Astronomy (SIfA), School of Physics, University of Sydney, NSW, Australia}
\altaffiltext{8}{Steward Observatory \& College of Optical Sciences, University of Arizona, Tucson, AZ, 85721, U.S.A.}
\altaffiltext{9}{Astrobiology Center of NINS, 2-21-1, Osawa, Mitaka, Tokyo, 181-8588, Japan}
\altaffiltext{10}{Observatoire de la Cote d'Azur, Boulevard de l'Observatoire, Nice, 06304, France}
\email{jovanovic.nem@gmail.com}

\begin{abstract}
Extreme adaptive optics systems are now in operation across the globe. These systems, capable of high order wavefront correction, deliver Strehl ratios of $\sim90\%$ in the near-infrared. Originally intended for the direct imaging of exoplanets, these systems are often equipped with advanced coronagraphs that suppress the on-axis-star, interferometers to calibrate wavefront errors, and low order wavefront sensors to stabilize any tip/tilt residuals to a degree never seen before. Such systems are well positioned to facilitate the detailed spectroscopic characterization of faint substellar companions at small angular separations from the host star. Additionally, the increased light concentration of the point-spread function and the unprecedented stability create opportunities in other fields of astronomy as well, including spectroscopy. With such Strehl ratios, efficient injection into single-mode fibers or photonic lanterns becomes possible. With diffraction-limited components feeding the instrument, calibrating a spectrograph's line profile becomes considerably easier, as modal noise or imperfect scrambling of the fiber output are no longer an issue. It also opens up the possibility of exploiting photonic technologies for their advanced functionalities, inherent replicability, and small, lightweight footprint to design and build future instrumentation. In this work, we outline how extreme adaptive optics systems will enable advanced photonic and diffraction-limited technologies to be exploited in spectrograph design and the impact it will have on spectroscopy. We illustrate that the precision of an instrument based on these technologies, with light injected from an efficient single-mode fiber feed would be entirely limited by the spectral content and stellar noise alone on cool stars and would be capable of achieving a radial velocity precision of several m/s; the level required for detecting an exo-Earth in the habitable zone of a nearby M-dwarf.  
\end{abstract}

\keywords{Astronomical Instrumentation, Instrumentation: adaptive optics, Instrumentation: spectrographs,  Extrasolar Planets}

\section{Introduction}\label{sec:intro}
Spectroscopy is a powerful technique widely used in astronomy. It was initially developed as a tool to determine the chemical composition and abundances of a target, and the first stellar spectrum was that of the sun taken by Fraunhofer in $1814$. Spectroscopy is now regularly used to study the kinematics of stellar systems, including their proper motion and orbital properties (in case of binarity or an exoplanetary system) and distance from Earth (i.e. redshift, in the case of distant galaxies). Low resolution spectroscopy has been used to map the structure of galaxies throughout the Universe~\citep{colless2001} and to confirm the presence of dark energy/matter and the expansion of the Universe~\citep{riess1998}. More locally, high-resolution spectroscopy has been used to understand stars and stellar structure by studying well-constrained systems like eclipsing binaries~\citep{las2002}. It has also been prolific in its detection yield of exoplanets~\citep{mayor95} and has been recently used for exoplanet characterization as well~\citep{snell2010,snell2014}. Spectroscopy provides a vast amount of information and is one of the key techniques employed throughout all areas of astronomy.

Classically, spectrographs were mounted at either the Cassegrain or Coude focus, and fed via a slit~\citep{huggins1895}. While having the instrument directly mounted to the telescope provides the highest throughput and is therefore still common for instruments intended for spectroscopy of dim objects (MOSFIRE~\citep{mclean2012}, for example), it has several shortcomings for high-resolution spectrographs, where stability is  often key. Instruments moving with the telescope (e.g. Cassegrain focus) experience a varying gravity vector and hence are prone to flexure. The Coude mounted version has the benefit of constant gravitational load and better insulation from environmental changes like temperature, as the Coude room can be isolated from the dome; however, beam transport requires multiple reflections and is somewhat lossy. Fiber fed instruments take this concept further by decoupling the spectrographs location from the telescope's foci. 

The advantages of feeding the instrument with a fiber instead of a slit were recognized early on, and in the late $1970$'s, the first fiber fed spectrographs were demonstrated~\citep{hill1980,hill1988}. Fibers provide an additional degree of decoupling in that they can be fed into a vacuum tank, where it is possible to realize a rigid fiber slit mounted directly to the spectrograph bench~\citep{quir2014}. Also, they provide what is termed scrambling (see Section~\ref{sec:enhspec}), which stabilizes the input illumination to the spectrograph optics~\citep{Stu2014}. 

Multimode fibers (MMF) with large cores and opening angles of the accepted light cone (quantified as the numerical aperture: the sine of the half-cone angle) were chosen for their superior collecting power from the seeing-limited telescopes of the day and have since been almost exclusively used to feed spectrographs. However, it was quickly realized that these fibers have several shortcomings. One issue was the fact that the focal ratio coming out of the fiber was degraded with respect to the injected focal ratio; this effect is known as focal ratio degradation (FRD)~\citep{ramsey1988}. With a larger cone angle at the output of the fiber, the spectrograph optics must be designed larger than desired, increasing the size and cost and reducing the stability of the instrument. The second issue was that the fibers did not fully scramble the image of the PSF incident at the input and indeed preserved significant spatial structure at the output. This resulted in a spatially and temporally varying slit illumination\footnote{We define the following terms here: the term PSF is used to describe the actual point-spread-function of the telescope including the atmosphere, which gives the light intensity distribution in the focal plane and is incident on the fiber input; the slit illumination (SI) or near field, refers to the input illumination of the spectrograph, either through a slit or at the exit of the coupling fiber (where the term near field comes from); the far-field illumination (FF) is the intensity distribution in the pupil of the spectrograph (far-field of a coupling fiber); and the spectral line spread function (SLSF~\citep{spro2012a, spro2012b}) describes the profile of a monochromatic line as seen by the spectrograph in its focal plane.}  structure being injected into the spectrograph, complicating calibration. To address this shortcoming, scrambling techniques to eliminate fluctuations in the spatial structure of the near and far fields have been developed which include using non-circular fibers~\citep{chaz2012,roy2014} as well as double-scrambling with ball lenses~\citep{hunter1992}. Also, modal noise (see below) limits the ultimate signal-to-noise ratio (SNR) achievable with a MMF-fed instrument to levels below what is required for high precision Doppler measurements or stellar spectroscopy~\citep{baud2001}. To reduce modal noise, fiber shakers have been developed which provide mechanical agitation~\citep{plav2013}. Although these scrambling techniques work well, they are not perfect and will limit the accuracy of SLSF calibration for ultra-high precision measurements.

Over the past $20$ years, adaptive optics systems have improved significantly and now diffraction-limited PSFs are commonplace at large observatories and are integral parts of future extremely large telescopes (ELTs). With a diffraction-limited telescope PSF, it becomes possible to utilize single-mode fibers (SMF) instead, which eliminate FRD and modal noise. We have recently demonstrated highly efficient coupling to SMF with the SCExAO instrument at Subaru Telescope~\citep{jovanovic2014,jovanovic2016a}. This opens the door to applying photonic technologies to efficient instrument design. With their compact footprint and broad functionality, photonic technologies will no doubt play a key role in future instrumentation. 

The aim of this work is to provide an overview of how the stable, high Strehl PSFs provided by advanced adaptive optics systems can enhance astronomical spectroscopy~\citep{Schwab2012}. This is an exciting topic that was recently discussed in a short article by~\cite{crepp14}. In Section~\ref{sec:ExAO}, we review the state-of-the-art in adaptive optics systems, and in Section~\ref{sec:enhspec}, we describe how these properties could be harnessed to develop innovative spectrograph concepts and enhance spectroscopy. Section~\ref{sec:future} details other innovative applications of adaptive optics for spectroscopy and Section~\ref{sec:app} outlines one possible instrument concept enabled by this technology. Section~\ref{sec:summary} rounds out the work with some concluding remarks.

\section{Advanced technologies for efficient fiber injection}\label{sec:ExAO}
\subsection{Adaptive optics - wavefront control}\label{sec:AO}
Over the past decade, adaptive optics (AO) systems have become widely available at larger ($5$-$11$~m class) observatories. Exploiting the improved light concentration delivered by such systems for spectroscopy seems like an obvious next step. Indeed, in the case of single object spectroscopy, the target star can serve as a natural guide star, and the required field-of-view is extremely small, both ideal conditions for running an AO system. In this section, we give a brief overview of the state-of-the-art in AO facilities and describe other key technologies that can also be exploited for the purposes of spectroscopy.  

AO systems consist of two primary components: a wavefront sensor to analyze the incoming wavefronts and a deformable mirror (DM) which can be used to correct the detected aberrations. The type of wavefront sensor used in an AO system will dictate the sensitivity to aberrations, while the number of sensors and deformable mirrors will determine the field-of-correction (i.e. the field over which the telescope's PSF is improved). In the simplest and most widely used incarnation, as a single-conjugate AO (SCAO) system, a single wavefront sensor senses aberrations produced by turbulence along a single line of sight and applies the cumulative correction to the DM, which is conjugated to a given altitude above the telescope. Due to line-of-sight effects, the measured aberrations are only valid over a small region known as the isoplanatic patch ($2-3$ arcseconds at $0.5~\mu$m for Mauna Kea), and the correction drops very quickly outside this field. Typical SCAO systems used at observatories like Keck and Subaru offer several hundred actuators over the area of the deformable mirror ($349$ for Keck~\citep{wiz2000} and $188$ for Subaru Telescope~\citep{minowa10}) and can achieve $30$-$40\%$ Strehl ratios in the H-band in median seeing. For clarity, the Strehl ratio is defined as the ratio of the peak flux of a measured PSF as compared to the ideal (aberration free) PSF for a given optical system and ranges from close to $0$ when the beam is heavily aberrated to $1$ for a for an unaberrated PSF.  SCAO systems use the target as a natural guide star which is ideal as this minimizes the difference in induced aberrations between the guide star and the target and hence offers the best correction. However, this limits high performance (high Strehl ratio) operation to targets brighter than $9-10^{th}$ magnitude in the sensing band. Below this level, the loop has to be run slower to accumulate enough photons to make an accurate estimation of the wavefront, which means that the loop can no longer keep up with the atmosphere and there is an associated drop in performance. This trend continues until the noise floor of the detector in the WFS is reached at about $\sim15-16^{th}$ magnitude, at which point the sensor no longer works~\citep{minowa10}.

Extreme AO systems can offer superior Strehl ratios by correcting for higher temporal and spatial frequencies of the atmospheric turbulence than traditional SCAO systems. To achieve this, ExAO systems must use wavefront sensors simultaneously offering high speed, high sensitivity and a large number of measured wavefront modes. To facilitate this, wavefront sensors utilize very sensitive cameras that have fast readouts and low latency, while the AO control loop commonly exploits FPGAs or GPUs for computations to maintain high loop speed. To correct for the higher order aberrations (smaller spatial scale of the turbulence), DMs with a larger number of actuators across the pupil than in standard SCAO systems are implemented. The number of actuators across the DM defines the control radius around the PSF in the focal plane~\citep{Opp2003} and hence the extent to which the light from the halo around the PSF can be reached and concentrated into the core of the PSF by the AO system. ExAO systems utilize DMs with several thousand actuators to achieve high Strehl ratios. 

There are currently six AO systems online that can achieve ExAO performance (Strehl ratio $\sim90\%$) and their properties are summarized in Table~\ref{tab:exao} and compared to a conventional SCAO system (Keck AO). It is clear that all ExAO systems currently in operation do indeed have many more illuminated actuators for correction. In addition, the loop speed for all systems is greater than for a typical SCAO system which is facilitated by running the computer as a real-time computer, with shielded cores and processes to prevent the OS interrupting the loop. This offers very low latency which is critical to fast operation. 

\begin{figure}
\centering 
\includegraphics[width=0.85\linewidth]{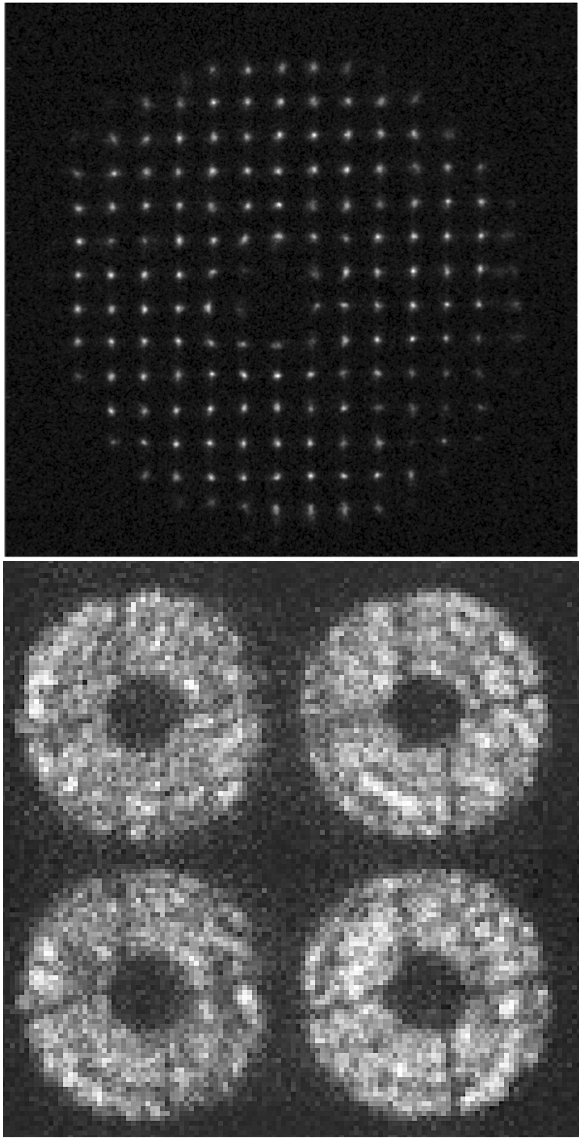}
\caption{\footnotesize (Top) An example Shack-Hartmann wavefront sensor image obtained on the $6.5$~m MMT. Credits: J. Codana \& O. Guyon, University of Arizona. (Bottom) An example PyWFS image obtained from the SCExAO instrument during closed-loop on-sky operation.} 
\label{fig:WFS}
\end{figure}

Shack-Hartmann and Pyramid wavefront sensors are equally used amongst ExAO systems. Shack-Hartmann wavefront sensors have a long history of use in astronomy and rely on sensing the local tilt in the wavefront by virtue of the position of the PSF formed by a micro-lens which subtends a sub-pupil. By measuring the position of an array of PSFs formed by a micro-lens array (shown in Fig.~\ref{fig:WFS}) which covers the entire telescope pupil, it is possible to reconstruct the wavefront across the pupil. To increase the sampling (i.e. number of micro-lenses across the pupil) as required to sense higher order aberrations, it is important to have very sensitive cameras. The existing generation of ExAO systems all utilize high quantum efficiency CCD-based cameras which have a read-out noise of $\sim1$ electron in typical operation conditions. 

\begin{deluxetable*}{lccccc}[h!]
\tabletypesize{\footnotesize}
\centering
\tablecaption{ExAO system characteristics compared to a conventional SCAO system (Keck AO). \label{tab:exao}}
\tablehead{
\colhead{Instrument}  	& 	\colhead{Telescope}	& 	\colhead{Wavefront sensor}  	&  \colhead{Detector}  	&    \colhead{Number of DM actuators}  	&   \colhead{Loop}        	\\  
\colhead{}  			& 	\colhead{}			& 	\colhead{type} 				&  \colhead{}  			&    \colhead{in the illuminated pupil}  	&   \colhead{Speed (Hz)}}  \\  
\startdata
GPI~\citep{macintosh14}	&  	Gemini-South		& 	Spatially filtered 			&    CCD				&   $2000$						&  $1000$				\\
					&					&	Shack-Hartmann			&					&								&					\\
					&					&							&					&								&					\\
SPHERE~\citep{bez08}	&  	VLT 				&   	Spatially filtered 			&    EMCCD			&   $1400$						&  $1200$				\\
					&					&	Shack-Hartmann			&					&								&					\\
					&					&							&					&								&					\\
P1640~\citep{dekany13}	&     Palomar			& 	Shack-Hartmann 			&    CCD				&   $2500$						&  $2000$				\\
					&					&							&					&								&					\\
					&					&							&					&								&					\\
LBTAO~\citep{esp11}	&      LBT				& 	Pyramid 					&    CCD				&   $672$							&  $1000$				\\ 
					&					&							&					&								&  ($2000$)			\\
					&					&							&					&								&					\\
MagAO~\citep{close13}	&      Magellan			& 	Pyramid 					&    CCD				&   $540$							&  $1000$ 			\\ 
					&					&							&					&								&  ($2000$)			\\
					&					&							&					&								&					\\
SCExAO~\citep{jov2015a}	&      Subaru			&	Pyramid 					&    EMCCD			&   $1600$						&  $3600$				\\ 
					&					&							&					&								&					\\
					&					&							&					&								&					\\
Keck AO~\citep{wiz2000}	&      Keck	II			&	Shack-Hartmann 			&    CCD				&   $240$							&  $670$				\\ 

\enddata
\tablecomments{CCD - Charged coupled device, EMCCD - Electron multiplying CCD. Values in parentheses are targets for future upgrades. SCExAO uses the commercially avialable OCAM$^{2}$K EMCCD from First Light Imaging.}
\end{deluxetable*}

The other wavefront sensor (WFS) choice for ExAO is the Pyramid wavefront sensor (PyWFS). The PyWFS combines high sensitivity to both low and high order aberrations and can be made linear over a wide range with tip-tilt modulation~\citep{guyon05}. The PyWFS utilizes a pyramid prism~\citep{esp10} positioned in the focal plane such that the tip of the pyramid subdivides the PSF and generates $4$ images of the pupil on the detector (shown in Fig.~\ref{fig:WFS}). Subdivision of the PSF at the focus results in phase modulations being converted to amplitude modulations that can be sensed by the detector. By sensing changes to the illumination of the pupil images, a PyWFS can be used to determine the phase errors entering the telescope; the same requirements for loop speed and camera sensitivity still apply. 

Shack-Hartmann WFSs were chosen by the early ExAO systems because of heritage and accumulated experience with these sensors. Newer PyWFSs are quickly becoming widely used because of their superior sensitivity to a given wavefront error~\citep{guyon05}. Manufacturing the pyramidal optic as a single element however is inherently difficult. Recently, we have demonstrated that this can be overcome by employing two roof prisms instead of a single pyramidal optic~\citep{jovanovic2016b}, making PyWFSs the approach of choice for future ExAO systems.

SCExAO at Subaru is the latest addition to the group of ExAO systems and a good case study for system architecture and hardware. It utilizes a camera system based on a deep depletion EMCCD which allows for high quantum efficiencies out to $950$~nm. In addition, the optimized read-out electronics of this system enable frame rates of up to $3.6$~kHz to be used for wavefront control for the first time using a commercially available camera. For correction it exploits a $2000$ element MEMs-based deformable mirror. Thanks to advances in these technologies over the past $10$ years, ExAO systems are now regularly delivering Strehl ratios of $>80\%$ in the near-IR in median seeing conditions.

It is worth noting at this point that ExAO systems are typically developed for the purpose of high contrast imaging at small angular separations from the host star. Thus, they are equipped with various coronagraphs, interferometers and low order wavefront sensors (LOWFS) that can also be exploited. As an example, the Lyot-based LOWFS (LLOWFS~\citep{singh14}) on SCExAO enables sub-milliarcsecond RMS control of the jitter in the position of the PSF~\citep{singh15}. This superior level of tip/tilt control is ideal for optimizing the injection into a fiber and maintaining the high stability pointing needed for long exposures for spectroscopy. 

\subsection{Pupil apodization}\label{sec:PA}
With ExAO systems delivering high Strehl ratios, and LOWFSs holding the PSF stable at the sub-milliarcsecond level, the light concentration and overall PSF quality is greatly improved, which makes it possible to efficiently inject into diffraction-limited photonic waveguides for the first time. Photonic technologies rely on guiding light only in the fundamental waveguide mode which has a Gaussian intensity distribution and a flat wavefront/phasefront. To efficiently couple light into such components, the properties of the incident beam must match these characteristics. The flat wavefronts provided by ExAO systems are ideal for this application. This has been exploited for the purposes of nulling interferometry on P$1640$ by~\cite{serebyn2010}. 
More recently, we have also explored this on the SCExAO instrument~\citep{jovanovic2014,jovanovic2016a}. In this work, we investigated matching the incident beam shape with that of the waveguide mode (i.e. a Gaussian) with the aid of apodization optics (known as Phase Induced Amplitude Apodization (PIAA) opitcs) that were used in the instrument for coronagraphy~\citep{guyon03}. These optics converted the flat-top pupil illumination, which induced diffraction rings in the focal plane (Airy rings), and limited the maximum coupling efficiency assuming a flat wavefront to $\sim80\%$~\citep{shak88}, into a near-Gaussian (prolate-spheroid) that has perfect overlap with the mode guided by the fiber. We demonstrated that the coupling efficiency into SMF with the apodization lenses was $>65\%$, when tested with an internal turbulence simulator and ExAO levels of wavefront correction ($90\%$ Strehl ratio, $80$~nm RMS wavefront error at $1550$~nm)~\citep{jovanovic2016a}. It was also shown that the coupling efficiency reduced linearly with decreasing Strehl ratio, offering a simple way to relate Strehl ratio to coupling for the purposes of planning future instrumentation. The majority of the coupling loss in the ExAO regime was due to residual wavefront error ($7\%$) and diffraction effects from the secondary support structures ($9\%$). This level of coupling into SMFs is comparable to those achieved with typical MMFs used for spectrographs at seeing-limited telescopes ($>70\%$ depending on the size of the seeing spot with respect of the fiber). This opens the possibility of exploiting photonic/diffraction-limited technologies for spectroscopy, which is the the focus of the following sections.

\section{Enhancing spectroscopy with adaptive optics}\label{sec:enhspec}
\subsection{The basic spectrograph equations}\label{sec:specequ}
The light concentration enabled by an AO system and the resulting possibility of feeding single-mode fibers or few-port photonic lanterns leads to several advantages for high-resolution spectrographs. The first is miniaturization. To understand this, we must look at the basic equations governing spectrograph performance. Equations~\ref{eq:resolution} and~\ref{eq:resolving_power} show the resolution ($\delta \lambda$) and resolving power ($R$) of a spectrograph\footnote{The terms resolution and resolving power are used loosely and interchangeably in astronomy. Here we stick to the original definitions.}, respectively~\citep{schro99}.
\begin{equation}\label{eq:resolution}
\delta \lambda=\frac{r\, \phi \, D_{\small{Tel.}}}{A\, d_{\small{col.}}}
\end{equation}
\begin{equation}\label{eq:resolving_power}
R = \frac{\lambda}{\delta \lambda} = \frac{\lambda A\, d_{\small{col.}}}{r\, \phi\, D_{\small{Tel.}}}
\end{equation}
The parameters in the equations are defined as follows: $\lambda$ is the wavelength of light, $r$ is the anamorphic factor, $\phi$ is the angular slit size, $D_{\small{Tel.}}$ is the diameter of the telescope, $A$ is the angular dispersion, and $d_{\small{col.}}$ is the diameter of the collimated beam incident on the disperser. It can be seen that the resolving power is inversely dependent on the slit size. Hence, the resolving power is maximized when the slit size is at a minimum (i.e. diffraction-limit). To maintain the same resolving power with a diffraction-limited feed, the collimator diameter must be decreased. This leads to proportionally smaller sizes for the other spectrograph elements, like the disperser and camera optics, and hence to miniaturization of the entire instrument. 

To illustrate this point, the diameter of a typical MMF used on facility spectrographs currently is between $100$ and $200~\mu$m, while a single-mode fiber has a core size of $\sim8~\mu$m (note both are used at similar focal ratios). This factor of $12.5$-$25$ in the ratio of slit sizes directly results in a corresponding reduction in the resolution of the MMF-fed system with respect to the SMF-fed system. To maintain the resolution of the MMF-fed system with a diffraction-limited feed, then the diameter of the collimating optic, disperser and camera optics can be reduced by this same factor, which can have tremendous impact on cost, spectrograph design and performance. As the MMF size required for the seeing-limited case scales linearly with telescope diameter, the advantage of using a SMF feed is especially amplified on larger telescopes, which otherwise require very large spectrograph optics. For example, the gratings for the HIRES spectrograph at Keck and the G-CLEF spectrograph for GMT are both $0.3\times1.2$~m in size~\citep{vogt1994, szent2014}; the cost of such an element is several hundred thousand dollars. 

\begin{figure}
\centering 
\includegraphics[width=0.95\linewidth]{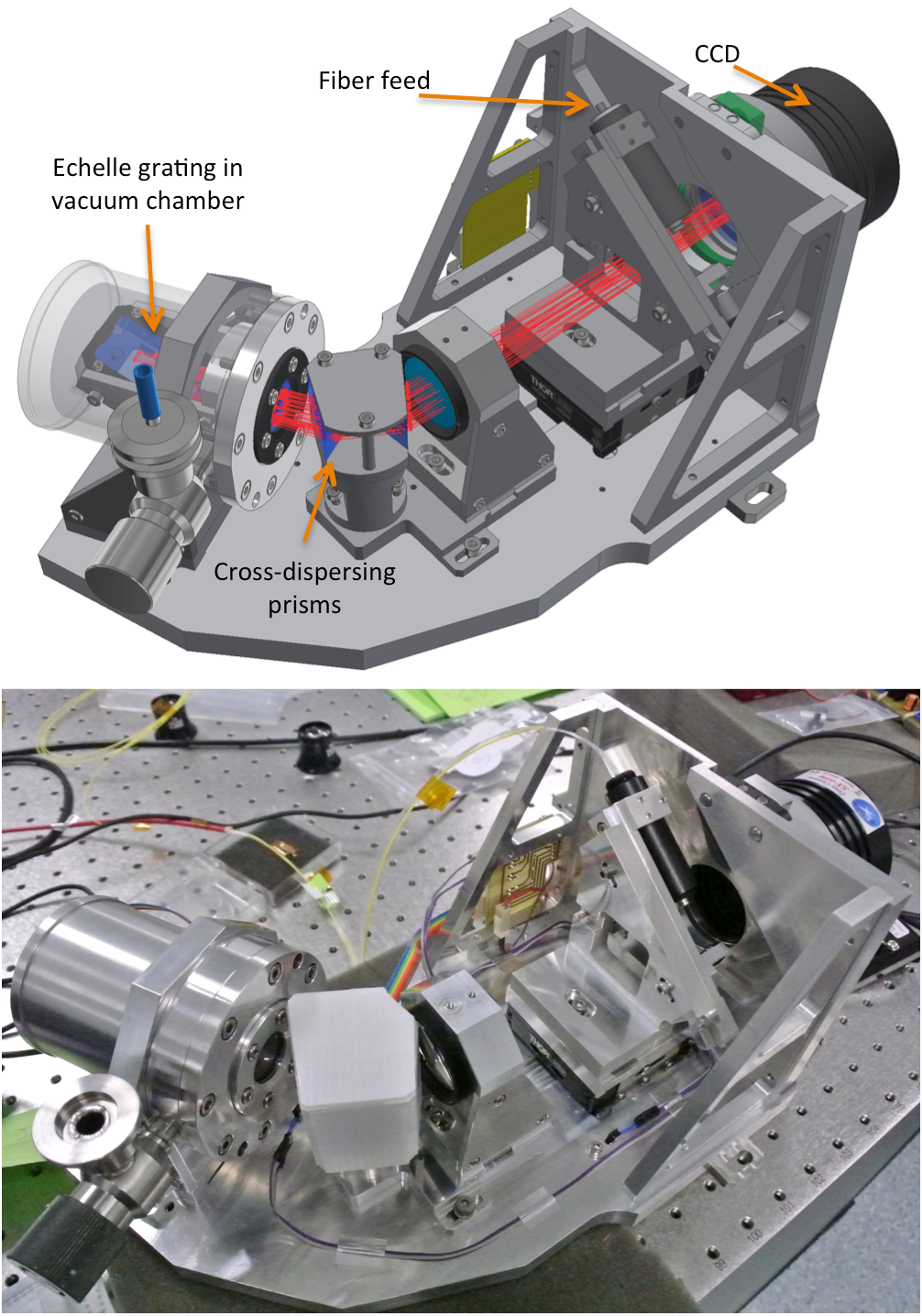}
\caption{\footnotesize An image of the RHEA spectrograph deployed at Subaru Telescope~\citep{feger2016,rains2016}. (Top) The 3D CAD rendering including the light rays (in red). (Bottom) An as-built image of the instrument. For a sense of scale, the instrument is sitting on a standard breadboard with $25$~mm hole spacing. Credit: T. Feger, Macquarie University.}
\label{fig:rhea}
\end{figure}

The direct impact from miniaturization when feeding a bulk optic-based spectrograph with a diffraction-limited feed include:
\begin{enumerate}
\item The smaller beam diameter enables the use of a simpler design with fewer optical elements, relaxed tolerances and a larger choice of materials. This results in higher throughput of the optical train inside the spectrograph and optimal resolution.
\item Large cost savings by using smaller components (optical and mechanical) that can often be off-the-shelf. 
\item The spectrograph fits into a much smaller envelope. This results in less mechanical flexure and more precise temperature and pressure stabilization, which translates to higher spectral stability and hence radial velocity precision.
\end{enumerate}
The Replicable High-resolution Exoplanet and Astroseismology spectrograph (RHEA) is the embodiment of all of these advantages (see Fig.~\ref{fig:rhea}). The spectrograph is fed by a single-mode fiber and is very small, occupying a footprint measuring $<30$~cm on a side, is built from off-the-shelf optics, and delivers a resolving power of $>70000$ from $430-670$~nm~\citep{feger2014a}. The R$2$ Echelle grating is easily stabilized in a vacuum chamber due to its compact profile ($5\times2.5$~cm in size), and with very simple temperature control, the spectrograph can achieve an intrinsic precision of $\sim1$~m/s over a time period of days~\citep{feger2016}. In addition, to maximize detector usage, it is possible to feed multiple (up to $10$) fibers into it at once. The relatively low cost of $\sim\$15$k makes it possible to replicate such simple spectrographs and deploy them widely.

In addition to the advantage of compactness, a diffraction-limited feed makes the spectrograph design independent of the telescope properties. To demonstrate this, the equation for the slit-size in the diffraction-limited case~\citep{schro99} can be expressed as 
\begin{equation}\label{diffraction}
\phi  \approx \frac{\lambda}{D_{\small{Tel.}}}.
\end{equation}
Substituting Eq.~\ref{diffraction} into Eq.~\ref{eq:resolution} and~\ref{eq:resolving_power}, we obtain
\begin{equation}\label{eq:resolutiond}
\delta \lambda=\frac{r \, \lambda}{A\, d_{\small{col.}}}
\end{equation}
and
\begin{equation}\label{eq:resolving_powerd}
R = \frac{\lambda}{\delta \lambda} = \frac{A\, d_{\small{col.}}}{r}.
\end{equation}
In this regime, it is clear that both parameters are now independent of the telescope diameter and are only a function of the choice of the optics in the spectrograph itself (i.e. the angular dispersion and collimated beam size). This means that a high-resolution spectrograph can be built without consideration of the telescope it will be used at and that this spectrograph would perform equally well at any observatory with a diffraction-limited feed (ignoring possible instabilities due to different environmental conditions such as temperature, pressure, etc). This advantage has also been exploited by the RHEA instrument. It was recently moved from the $0.4$~m telescope at Macquarie University Observatory, where the early prototyping and science was done, to the $8.2$~m Subaru Telescope in Hawaii~\citep{rains2016}. Due to the SMF-feed, the instrument has the same resolving power at both telescopes. 

The second advantage of being able to couple into SMFs is the very high level of control and knowledge of the input illumination/wavefront entering the spectrograph, which cannot be achieved with conventional MMFs and even less so with a slit. SMFs act as modal filters, transmitting light only in the fundamental waveguide mode which is characterized by a Gaussian intensity distribution and a flat wavefront. This property has been exploited in stellar interferometry for $20$ years since it allows for robust, high-contrast fringes to be realized and has been termed ``spatial filtering"~\citep{Coude1994}. Spatial filtering offers advantages for spectroscopy as well. 

Currently, the major limitation to the performance of spectrographs is associated with the preservation of spatial structure from the input beam to the output of the fiber. In an ideal instrument, the output spatial distribution of a fiber would be constant and independent of the input conditions and of perturbations to the waveguide. Thus the SLSF would be invariant with time and easy to calibrate. In practice MMFs retain some of the information of the spatial distribution from the input because they guide light in a subset of modes (not a continuum). In an ideal waveguide, the power distribution amongst the subset of modes would be preserved as the light propagated along the waveguide, but in practice the power redistributes between modes and can even excite new modes. This effect is called modal noise and is due to an imperfectly fabricated waveguide (core size and index of refraction are not uniform along the length of the fiber, for example) and to bends, stress, strain and other environmental factors. In addition, the input beam is typically changing with time due to a fluctuations in the AO system or seeing (for a telescope that does not have an AO system). In short, if the near-field and far-field of the waveguide are not evenly illuminated at the input and modal noise is present, the output beam that is imaged onto the detector will vary with time and this will lead to the misinterpretation of Doppler shifts in the spectrum of the target star. To overcome this problem, a lot of effort has gone into the investigation of non-circular fibers~\citep{chaz2012,roy2014} and double scramblers~\citep{hunter1992,hal2015a}, which aim to make the fiber feed output independent of the input. Modal noise is commonly addressed by rapidly agitating the fiber. Agitation continually redistributes the power among the available modes, so that the modal distribution gets averaged over the exposure time~\citep{plav2013}. While a lot of progress has been made, the resulting scrambling performance is not perfect. SMFs, on the other hand, do not require scrambling or agitation; they transport one mode only, and the output beam is completely independent of the input coupling conditions, which instead manifest only as intensity fluctuations. 

In addition, the light source used for calibration (i.e. a laser frequency comb, hollow cathode emission lamp, stabilized etalon) or gas cell should produce the same illumination as the science light. Laser frequency combs, which offer the highest long-term calibration precision, are inherently diffraction-limited and when used on MMF-fed spectrographs, they suffer from modal noise and insufficient scrambling. This is problematic since one cannot be sure that the illumination of the calibration light is identical to that of the star and stable with time. As a result, scrambling techniques and ways to decorrelate the diffraction-limited output of the calibrator have to be employed~\citep{mah14}. In a SMF-fed spectrograph, any calibrator coupled to the SMF has the same outgoing illumination and wavefront as the starlight, by definition. By eliminating the effects of scrambling and modal noise on both the science and calibration fiber, SMF feeds will improve the precision with which a high-resolution spectrograph can be calibrated and enhance precision radial velocity measurements.  

The iLocater spectrograph~\citep{crepp16} at the LBT will be the first facility instrument purpose built to employ a singe-mode fiber feed behind an ExAO system. It was specifically designed to exploit the advantages outlined in this section. It will achieve a $R=150000-240000$ and operate in the y and J-bands. Preliminary tests of a prototype fiber injection, which does not exploit pupil apodization, are yielding promising coupling efficiencies of up to $25\%$ in y-band with further improvements to come for the final instrument~\citep{bechter2016}. This instrument shows that a shift in the approach to instrument design is underway and will be a very important yardstick for future instrumentation concepts.  

This brings us to the third advantage of being able to couple into SMFs: the ability to exploit photonic technologies and functionalities. Here we define a photonic component as a device that guides light, whether it be an optical fiber or a structure on a chip. The development of photonic technology throughout the past $30$ years has been driven by the multi-billion dollar telecommunications industry. With this level of investment, many devices spanning a wide range of functionalities have been realized and are now readily available. In the next section we review some of the key components/functionalities that could be utilized for enhancing spectroscopy. 

\subsection{Photonic functionalities}\label{sec:photonics}
\subsubsection{Mode conversion}
Photonic lanterns are devices that can be used to convert multimode beams into single-mode (diffraction-limited) beams~\citep{saval2013}. In its original format, the device consisted of several SMFs bundled together and tapered down to form a single MMF. This device was first realized in $2005$~\citep{saval2005} and intended to be used to capture seeing-limited light with the MMF end of the device from the focal plane of a telescope. However, to efficiently convert between the MMF end and the SMF end, the number of modes excited in the MMF end has to be matched by the number of SMF output ports. The number of modes in a telescope PSF can be approximated by the following equation
\begin{equation}\label{eq:mode_num}
N_{m} =\frac{\pi^{2}}{16}\left(\frac{\theta_{seeing}D_{Tel.}}{\lambda}\right)^{2}
\end{equation}
where $\theta_{seeing}$ is the seeing in radians, $D_{Tel.}$ is the diameter of the telescope in meters and $\lambda$ is the wavelength in meters~\citep{spal2013}. If we take the case of the Australian Astronomical Telescope (AAT) as an example, where the diameter is $3.9$~m and the seeing is $\sim 2$ arcsec we find that the PSF supports $350+$ modes at $1.55~\mu$m. To efficiently capture this light, the device needs to have $100$'s of SMFs~\citep{noord2009}, reducing the average flux in each output fiber by the corresponding factor. From equation~\ref{eq:mode_num} it is clear that on smaller telescopes ($<0.7$~m in diameter), it is possible to subdivide the PSF into $<10$ fibers efficiently at the same site. This is a result of the telescope diameter approaching the Fried parameter, $r_{0}$ for the site, and the telescope no longer being limited by the seeing but by diffraction. Hence, most of the light becomes concentrated in a single core in the PSF and can be efficiently collected by only a few mode device. 

As one moves towards shorter wavelengths (e.g. visible part of the spectrum), such devices would require several thousand ports, which is beyond current technology. For this reason the integration of this device onto a single monolithic chip has been investigated. \cite{thomson2011} demonstrated the first integrated device fabricated via ultrafast-laser inscription whereby a laser with femtosecond length pulses is focused into a glass and used to create a waveguiding region. The waveguide routes that form the device were sculpted by careful translation of the laser's focus throughout the bulk of the glass. Further optimization of integrated devices has been demonstrated and now devices with less than $5\%$ transition loss can be readily fabricated (note that total throughput is $\sim75-80\%$)~\citep{jovanovic2012,spal2013}. More recently, such devices have been tested on-sky behind AO systems as well~\citep{mac2014,harris2015}.

Another technological route to cope with the large numbers of waveguides needed are so-called multicore fibers, which combine many single-mode cores in a common cladding volume and are manufactured with regular fiber drawing methods. These fibers can also be tapered down to form photonic lanterns~\citep{birks2012}. Such a device has been recently tested on-sky to reformat the light into a pseudo-slit with a $50\%$ transmission efficiency (including coupling) behind the Canary AO system at the William Herschel Telescope~\citep{maclachlan15}.

Few-port photonic lanterns, such as $7$-port devices which consist of a central core surrounded by a single ring of six other cores, can bridge the gap for observatories which have AO facilities but not ExAO performance. Lower levels of wavefront correction and pointing precision, which would typically be inefficient to use for direct SMF injection, can still be utilized with a few-port lantern. With its larger physical size, the few port lantern is less sensitive to wavefront and pointing errors than a SMF, making it ideal for offering diffraction-limited performance when an ExAO system is either not available or the performance is limited (due to bad seeing, for example). In this way, this device enables photonic/diffraction-limited possibilities by relaxing constraints on the AO system. Such devices could also be used at small seeing-limited observatories that wish to exploit the advantages of a diffraction-limited spectrograph without the crippling losses associated with coupling directly into a SMF in such conditions~\citep{betters2013}. 

\subsubsection{Spectral filtering}
Fiber Bragg gratings (FBGs) offer the ability to spectrally filter light. They are versatile and can be tailored to remove light from parts of a spectrum with custom bandwidths and attenuations. These devices were explored for the removal of OH-lines which are caused by the recombination of molecules in the atmosphere at night. The FBGs fabricated for this application were among the most complex spectral filters ever realized and attenuated over $100$ lines across the H-band, with varying bandwidths and levels of attenuation~\citep{jbh2004, trinh2013}. These FBGs were utilized in the GNOSIS instrument to eliminate the OH line contamination in the spectrum of redshifted galaxies prior to injection into a spectrograph. This prevents the light of the bright OH lines from being scattered by the spectrograph optics and raising the background level across the spectrum~\citep{jbh2011}. In addition, FBGs could be used to convert the signal from a spectral shift in a target into an intensity fluctuation which can be more readily detected with fast avalanche photodiodes and utilized in time domain astrophysics~\citep{marien12}. Bragg gratings have also been integrated onto monolithic chips to open up the possibility for compactness and scaling in future~\citep{spal2014}.

\subsubsection{Spectroscopy}
\begin{figure}
\centering 
\includegraphics[width=0.95\linewidth]{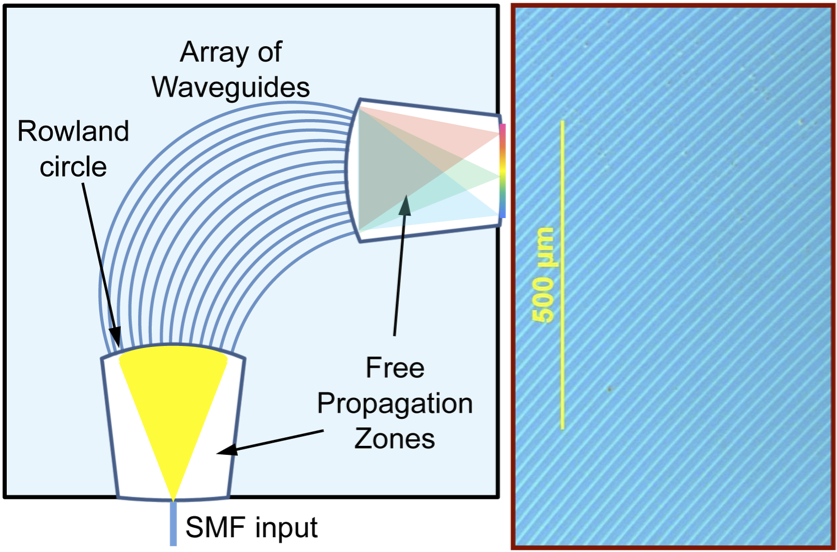}
\caption{\footnotesize (Left) Schematic of the key components of an AWG. (Right) Microscope image of the a section of the array of waveguides for a typical AWG device.}
\label{fig:AWG}
\end{figure}

In addition to devices for enhancing coupling at the telescope focus or for spectral filtering, entire spectrographs can be made from a single monolithic chip. Arrayed waveguide gratings (AWGs) are the most mature technology and have been explored for astronomical applications. AWGs are planar photonic devices which consist of $3$ sections: an input free-propagation zone (FPZ), an array of single-mode waveguides, and an output free propagation zone (shown in Fig.~\ref{fig:AWG}). The FPZs are essentially slab waveguides which are thin enough so as to guide light in a single-mode in the vertical plane while their width is such that they are multi-mode in the substrate plane. Light from a SMF is injected into the input FPZ. Although confined in the vertical direction, the light diverges in the substrate plane as it propagates along the FPZ. At the far end of the FPZ, an array of single-mode waveguides are uniformly positioned along what is known as the Rowland circle. This circle follows the curvature of a wavefront at the output of the FPZ, assuming a diffraction-limited injection at the input. In essence, the closely spaced SM guides collect the light along a single phase front of the diffracting wave. The guides route the light to a second FPZ. Each guide is longer than the previous one by a fixed amount. This creates a fixed optical delay between each guide so that when the beams are combined in the output FPZ, interference takes place and a spectrum is formed at the output of the device. In this way, the guides in the array act like the grooves of a diffraction grating and the entire device behaves like a phased array. 

Initial tests on AWGs for astronomical purposes focused on telecommunications-grade devices that could be easily sourced. These devices were optimized for operation around $1550$~nm, offered a free-spectral range (FSR) of $50$~nm, and resulted in a resolution of $R = 7000$. \cite{cvetojevic2009} first used the device to generate a full H-band spectrum of night time airglow by pointing the collection SMF towards the sky (no optics or telescope) and using a cross-disperser to unravel the orders. They also showed that it was possible to optimize the performance of an AWG so that signals from multiple SMFs could be injected simultaneously into a single chip and recovered via cross-dispersion~\citep{cvetojevic2012a}. This concept was tested on-telescope at the Anglo-Australian Telescope (AAT) at Siding Springs Observatory, behind the photonic lantern fiber-feed used for the GNOSIS instrument~\citep{cvetojevic2012b}. Although the throughput of the prototype was poor, future optimization will enhance the efficiency of photonic instrumentation. The tests were however, successful in demonstrating the potential for building compact instrumentation. The real advantage of AWG-based spectrographs becomes apparent when only a single free-spectral range is used so that the bulk optics of the cross-disperser can be eliminated and the detector butted up against the output of the chip. One could consider using a linear detector array, which is a more cost effective solution in this case, or stacking multiple AWGs on top of one another to enable multi-object capabilities on a $2$D detector as proposed by~\cite{jbh2010}.

\begin{figure}
\centering 
\includegraphics[width=0.8\linewidth]{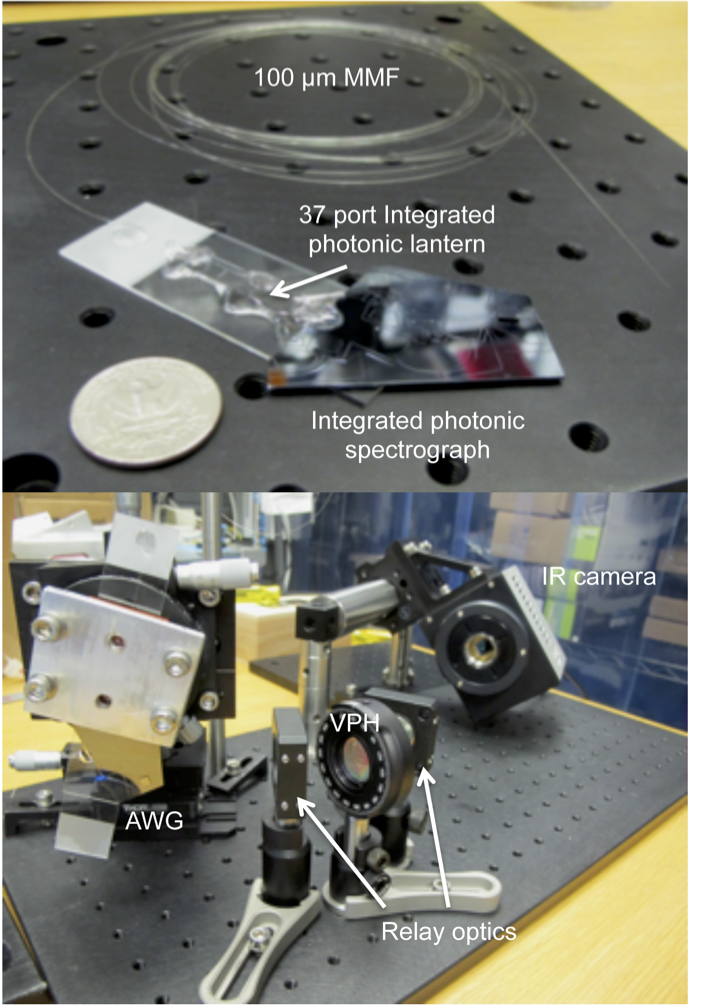}
\caption{\footnotesize (Top) An AWG directly bonded to an integrated photonic lantern. (Bottom) AWG in a low resolution cross-dispersed setup. VPH - Volume Phase Hologram}
\label{fig:AWGIPL}
\end{figure}

More recently, \cite{cvetojevic2016} have successfully combined an integrated photonic lantern with an AWG to make an all-photonic device. The device, depicted in Fig.~\ref{fig:AWGIPL}, was fed by a MMF and light injected from SCExAO without the ExAO capability running, ($30$-$40$\% Strehl ratio in the H-band from the Subaru AO system).  
The device consisted of a $37$-port lantern which was remapped into a $37$ waveguide long, linear slit at the entrance to the AWG. The entire cross-dispersed instrument offered $R=2000$ and is shown in the bottom of Fig.~\ref{fig:AWGIPL}, demonstrating that compact instruments can indeed be realized. This demonstration was particularly interesting because it stumbled upon a new type of modal noise which manifests because of the mismatch between the modes supported by the delivery MM fiber and the MM waveguide section at the input to the integrated photonic lantern. Although, not an ideal outcome, this work shows that photonic instrumentation can present unforeseen effects and we encourage the eager reader to view the full results elsewhere~\citep{cvetojevic2016}. In fact, at extreme precision radial velocity levels, a SMF is not truly single-moded either. A SMF supports $2$ orthogonally polarized modes. The non-zero birefringence of SMFs (typically $10^{-6}$) will break the degeneracy between the polarized modes, which will be seen as an illumination change on an Echelle grating (a polarization sensitive optic) and in this way introduce another subtle form of modal noise~\citep{hal2015b}. 

To get around the noise term observed with the previous photonic spectrograph and to take full advantage of the ever increasing Strehl ratio offered by the SCExAO instrument throughout commissioning, we developed and installed a SMF-fed AWG-based spectrograph. As this spectrograph was fed by a single fiber, a smaller amount of cross-dispersion was sufficient and was obtained from a prism. The total throughput of the spectrograph for unpolarized light was $44\pm3\%$ and it offered a resolving power of $\sim5000$ across the J and H bands~\citep{jovanovic2016a}. Preliminary coupling efficiency measurements were as high as $50\%$ at $1.55~\mu$m using the fiber injection described above in SCExAO. This demonstration shows that photonic prototypes are entering the regime of efficiency and performance where they can be competitive against traditional instrument designs. 

Much like a diffraction grating, it is possible to tailor the dispersion and control typical spectrograph properties such as the order of diffraction, the FSR, the central operating wavelength and so on by optimizing the parameters of the AWG. Spectrographs with astronomically relevant properties have been designed recently~\citep{cvetojevic2012c,cvetojevic2014}. These devices have been tailored for high-resolution spectroscopy in the near-infrared (NIR). Specifications achieved in the design include: $R=65000$ in the H-band, a theoretical average throughput of $62\%$ across the H-band, no polarization dependence, and a FSR of $16$~nm. With such a narrow FSR, the AWG is designed to operate across the H-band by cross-dispersing the output onto a HAWAII$2$RG array. The entire device would measure $\sim100\times50\times1$~mm, which makes thermal stabilization simple. This device is currently being fabricated. 

\subsubsection{Calibration}
\begin{figure}
\centering 
\includegraphics[width=0.99\linewidth]{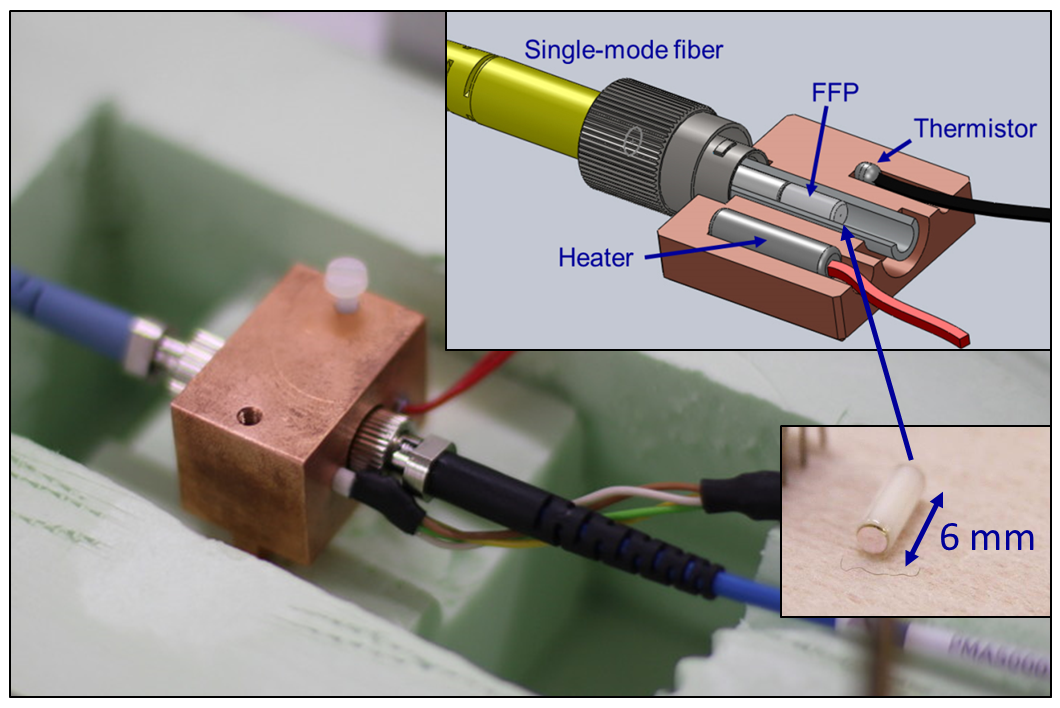}
\caption{\footnotesize An image of a compact, SMF-based Fiber-Fabry Perot etalon for precision calibration~\citep{gure14}.}
\label{fig:FFP}
\end{figure}

Finally, photonics offers the ability to make compact, simple and inexpensive calibrators for spectroscopy. As discussed previously, laser frequency combs (photonic devices themselves, of course) offer the ultimate in long-term stability and precision, at the expense of complexity and cost. However, if ultimate theoretical precision and knowledge of the absolute wavelength of each comb line is not required, then several other technologies become applicable. These technologies revolve around using SMF-based interferometers to create a comb-like spectrum for calibration. One version uses two simple fiber splitters/couplers to form a Mach-Zehnder interferometer~\citep{feger2014b}. Another is based on a Fabry-Perot etalon being established in a very short length of SMF by using mirror coatings on the ends~\citep{hal2014} (shown in Fig.~\ref{fig:FFP}). They both deliver a comb of lines at the output when a broadband light source is injected into the input, the spacing of which can be optimized by modifying the physical lengths of the devices. As demonstrated by~\cite{gure14} and \cite{schwab15}, the etalon version can be locked to an atomic transition as a wavelength reference, resulting in long term stability at the cm/s level. 

Integrated optic-based calibration can be obtained by exploiting ring-resonators~\citep{ellis2012}. These devices consist of a waveguide that transmits the broadband light from the calibration source which is placed in close proximity to a waveguide that forms a closed circle. In this architecture, light is only evanescently coupled to the circular guide if it is resonant with the guide. This resonance occurs at numerous discrete and periodic frequencies across the broadband spectrum, resulting in a spectrally filtered output. In this way, a comb could be generated from a miniature photonic chip. 

\section{Other applications of ExAO in spectroscopy}\label{sec:future}
It is clear that ExAO systems can offer numerous advantages in the design of future spectrographs as well as enable the use of numerous photonics functionalities. In addition, there are a number of ways they could be exploited to either enhance current spectroscopic techniques and/or enable entirely new areas of research. 

\subsubsection{Spatially resolved spectroscopy}
One obvious application is spatially resolved spectroscopy whereby components/structures of a binary, planetary system or disk are spatially resolved and the components spectroscopically studied individually. By selectively positioning the collection SMF over one of the two stars at a time, the star light contamination from the companion can be rejected more efficiently and hence a cleaner spectrum can be obtained for each component. For close-separation binaries, this is currently only possible in eclipsing systems where one star is blocked by the other at some point in the orbit. This could, for example, be used to investigate whether followup candidates of the Kepler, PLATO or TESS space telescopes are background stars or are part of the target system in a manner similar to that already undertaken by RoboAO's direct imaging followups~\citep{law2014,bara2016,zieg2016}. Due to the limited field-of-view of the SMFs, sky-contamination from excited OH lines in the NIR is $2$-$3$ orders of magnitude lower with a SMF than with a MMF feed.

In addition to individual fibers, it is also interesting to consider the use of a SMF-fed integral field unit (IFU). This would generally consist of a bundle of single mode fibers bonded to micro-lenses, forming a small optical element which subtends a part of the focal plane, segmenting the image and channelling the light from different parts of the image to the spectrograph via fibers. Indeed, the RHEA spectrograph at Subaru Telescope does just this~\citep{rains2016}. RHEA uses a $3\times3$ fiber IFU. The fibers are spaced by $1$~mm in the focal plane which corresponds to a $16$~mas spacing projected on-sky. The instrument was optimized for operation from $600$-$800$~nm at a resolving power of $\sim60000$. With its high spatial resolution, this instrument will be used to spectroscopically map the surfaces of giant stars in order to study convection and stellar rotation for the first time, study resolved protoplanetary disk structures as well as for the detailed characterization of accreting protoplanets. The final topic is extremely exciting with a recent discovery of an accreting exoplanet~\citep{sallum15}. An instrument like RHEA behind SCExAO could potentially shed light on the accretion rate and in-flow velocity, which would allow the mass and density of the forming exoplanet to be constrained. This data combined with the spatial location in the protoplanetary disk will aid in refining planetary formation models which are currently unconstrained. Given the reasonable cost and relatively small footprint of an instrument like RHEA, it is possible to envision an IFU which consists of $50\times$ as many elements tiling the entire focal plane feeding $10$'s of small spectrographs in an instrument optimized for exoplanet detection on an extremely large telescope. 

\subsubsection{Post-coronagraphic spectroscopy}
\cite{snell2010,snell2014} demonstrated a technique to extract high-resolution spectra from substellar companions. The technique relies on collecting a transmission spectrum which has the combined signal of the host and planet as it transits the host. After careful removal of telluric contamination from the Earth's atmosphere, they showed that it was possible to cross-correlate the spectra with a template spectrum for various molecular species such as carbon monoxide and methane and recover very low SNR detections of the former specie in the atmosphere of gas giants. Because of the associated Doppler shift during the period of the observation, it was also possible to tightly constrain the masses of the host and companion and other orbital parameters. This elegant technique demonstrated the ability to extract high-resolution spectroscopic information from substellar companions for the first time, albeit at very low SNR.  

One avenue to improving the SNR is by first suppressing the starlight with a coronagraph in order to remove the stellar noise, and then integrating on the partially isolated signal of the companion for longer. As outlined, most ExAO facilities are equipped with state-of-the-art coronagraphs to suppress the on-axis starlight.  By collecting the light from the planet and the residual flux from the star post-coronagraph, it is possible to reduce the photon noise by orders of magnitude, improving the SNR of the detected spectral features. If a SMF, or a MMF who's core is matched to the approximate size of the companion ($\lambda/D$) in the focal plane is used, this would add a second layer of rejection to contamination by the residual stellar flux. This technique is called post-coronagraphic high-resolution spectroscopy~\citep{kaw2014} and has thus far not been demonstrated in practice but is receiving growing attention. This is a promising method that will be explored with the combination of a fiber injection within SCExAO and the  Infrared Doppler (IRD) instrument~\citep{tamura2012} soon to be commissioned at Subaru Telescope. Again, if detector real estate were not a limitation, one could consider using an IFU for blind searches for new planets to speed up the detection process as well. 

\subsubsection{Visible light fiber injection}
With ExAO systems capable of $>90\%$ Strehls in the NIR, an improvement in the visible is also seen. Indeed, MagAO has demonstrated Strehls of $\sim50\%$ in the i'-band~\citep{males14a} ($35\%$ in the i'-band~\citep{males14b}). This opens up the possibility of feeding visible light spectrographs with ExAO systems as well. With lower coupling efficiencies in this part of the spectrum, few-port photonic lanterns could play a larger role in optimizing light collection power. Despite the lower coupling efficiencies, the advantages of a single-mode feed (elimination of problems with imperfect scrambling and modal noise, miniaturization of the spectrograph, and the ability to use photonic devices) make such a instrument well worth considering.

\subsubsection{Low-order wavefront sensing}
Photonic lanterns themselves could be used to sense drifts in the pointing and low order aberrations at the collection plane. In the case of ExAO correction with a high level of PSF stability, most of the light can be coupled directly to the central core of a few port lantern, which has the best overlap with the incident beam. Indeed, localizing the flux to a single fiber is desirable since it will result in the highest SNR and shortest exposure times. With low signals in the surrounding fibers, it may be possible to use those fibers in conjunction with avalanche photodiodes (for example) to monitor the PSF quality on fast time scales. Further, this information could potentially be used to drive a closed-loop correction of the modes by the DM. Note that the lantern's sensitivity to low order modes will be different from mode to mode and heavily dependent on the physical properties of the lantern (number of cores, size of MM end compared to beam size, core geometry), the injected focal ratio, and the amplitude of the wavefront error. Nonetheless, this method should at least be applicable to sensing tip/tilt. In this way, the photonic lantern used to reformat and collect the light would also become a LOWFS. The benefit of this over other types of sensors is that it would detect the residual wavefront errors in the plane where the injection happens, eliminating non-common path and chromatic errors. This could be a powerful tool in future instrument design. 

\subsubsection{Autonomous spectroscopy}
With the recent successful demonstration of autonomous AO systems for direct imaging by the Robo-AO project~\citep{law2014,bara2016,zieg2016}, one could consider using such a system to feed a compact photonic spectrograph. While this would not involve ExAO performance, high Strehl ratios could be achieved on smaller telescopes ($<2$~m), maintaining reasonable coupling efficiencies ($\sim30\%$). The compact and easy to stabilize nature of photonic spectrographs lends itself well to a low maintenance scheme such as this. In future, large spectroscopic surveys could be realized in this fashion. 

\section{Instrument concept}\label{sec:app}
There are numerous applications for the technology outlined above. Here we focus on one key example, namely ultra-high precision spectroscopy for the detection of Earth-sized exoplanets around M-type stars. The motivation for investigating Earth-sized analogs around such cool stars stems from the fact that the precision required to detect small companions around small stars is lower (several m/s compared to cm/s for G-stars), the orbital period for objects in the habitable zone (HZ) is shorter, making them quicker to find and confirm, and recent statistics from the Kepler candidates suggest that $15\%$ of cool stars have Earth-sized planets orbiting in the HZ~\citep{dressing13}. In addition, cooler stars have an abundance of molecular lines in the NIR where AO systems achieve a higher level of performance. Finally, the majority of stars in the solar neighborhood are cool stars, presenting a significant sample of targets. Indeed, it is for these reasons that instruments such as IRD~\citep{tamura2012}, Habitable zone Planet Finder (HPF)~\citep{mah2014}, SPIRou~\citep{et14} and CARMENES~\citep{quir2014} are currently being developed. 

The instrument concept comprises the following aspects: the light is to be injected from SCExAO, lossless apodization optics are to be used to improve coupling, a SMF is used to deliver the light to an AWG based-spectrograph, cross-dispersion is provided by an efficient volume phase hologram (VPH) and the detector is a HAWAII$2$RG (H$2$RG, $2$k$\times2$k pixels). The AWG used for this concept is the one described in Section~\ref{sec:enhspec} above ($R=65000$, $FSR=15$~nm). 

\begin{figure}
\centering 
\includegraphics[width=0.99\linewidth]{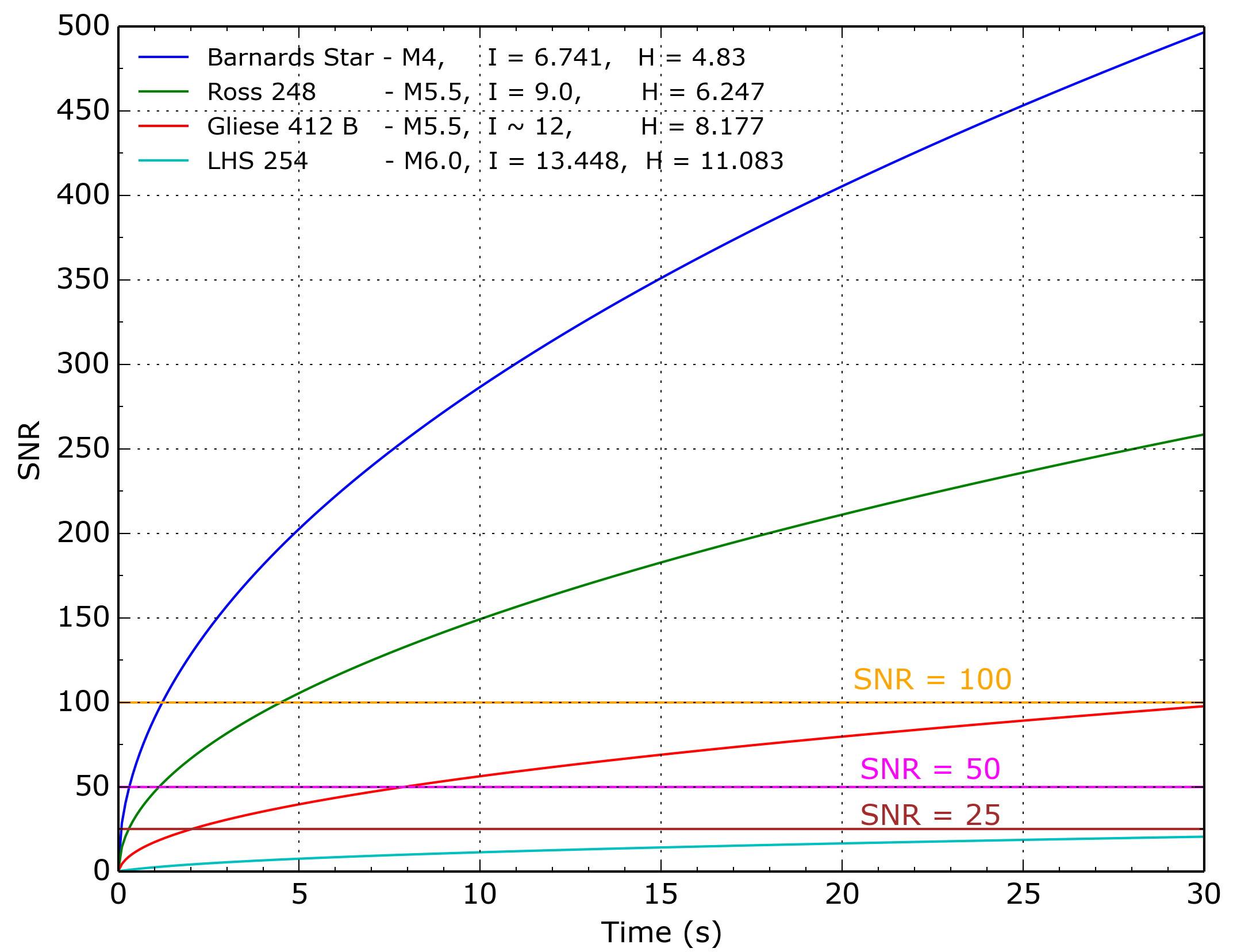}
\caption{\footnotesize The SNR per pixel for four M-stars with an AWG-based high-resolution spectrograph fed by the single-mode fiber injection in SCExAO. }
\label{fig:SNR}
\end{figure}

To estimate the performance of such an instrument, the signal-to-noise ratio (SNR) was calculated. The M-stars used in the simulation were chosen to cover a range of brightnesses from H-mag$=4.83$--$11.08$ in order to take into account the associated drop in AO performance at the faint end of the spectrum (see Fig.~\ref{fig:SNR} for star details). The estimated efficiency of such an instrument in the H-band is summarized in Table~\ref{tab:throughput}. With realistic figures for the throughput of all the elements of a diffraction-limited spectrograph behind SCExAO, the total system efficiency is $10$--$17\%$. The range in efficiencies is associated with the diminishing performance of the AO-correction for fainter guide stars.
\begin{deluxetable}{lc}[h!]
\tabletypesize{\footnotesize}
\tablecaption{Estimated throughputs of an AWG-based high-resolution, SMF-fed spectrograph on Subaru Telescope in the H-band.}
\centering
\tablehead{
\colhead{Element}	& 	\colhead{Throughput (\%)}} 	        \\ 
\startdata    
Atmosphere		& 	$97$							\\
Telescope	optics	& 	$92$							\\
AO$188$	optics	&	$73$							\\
SCExAO optics		& 	$78$							\\
Apodizer optics		&   	$95$							\\     
Fiber coupling		&   	$65 (90)$, $55 (75)$, $37 (50)$		\\     
SMF to AWG		&   	$97$							\\ 
AWG			&   	$62$							\\ 
Relay optics		&	$98$							\\
VPH				&	$92$							\\
H2RG QE			&	$68$							\\
\textbf{Total}		&	\textbf{$17$, $14$, $10$}			\\
\enddata
\tablecomments{QE- Quantum efficiency. Total includes all elements except the QE of the detector. Values taken from~\cite{jov2015a}. Values in parentheses reflect the assumed Strehl ratio for that level of coupling.
\label{tab:throughput}}
\end{deluxetable}

The coupling efficiency was determined from recent laboratory data collected on the SCExAO instrument~\citep{jovanovic2016a} and applied based on the brightness of the star in the wavefront sensing band (I-band for SCExAO). The experiment involved measuring the coupling efficiency of the fiber injection in SCExAO with apodization optics and a turbulence simulator running. In this way the coupling efficiency vs Strehl ratio and wavefront error were empirically determined for the SCExAO instrument. For I-mag $<9$, ExAO performance can be achieved ($90\%$ Strehl in the H-band, $\sim80$~nm RMS wavefront error). For I-mag $>9$, the Strehl ratio rolls off. The Strehl ratio vs star magnitude curve has not been characterized for SCExAO yet so we use the curve for the SAXO AO system~\citep{sauvage2016} (AO system of SPHERE) as a proxy to estimate the Strehl ratio. The laboratory determined coupling efficiency corresponding to a given Strehl ratio was used. All coupling efficiencies used for the simulation are summarized in Table~\ref{tab:throughput} and represent a best case performance in median seeing conditions. In reality the performance of the AO system on any given night could be reduced due to poor seeing or improper tuning which would cause a drop in the Strehl ratio by a factor of $2-3$ from those reported in table~\ref{tab:throughput}.  

The collecting power was assumed to be $53$~m$^{2}$ (corresponding to the $8$~m primary mirror at Subaru Telescope), and the spectra were Nyquist sampled ($2$ pixels for each spectral channel, $R=65000$, $\delta\lambda=0.025$~nm). Noise terms used in the calculation include photon noise, sky and telescope background, dark current ($0.046e^{-}pixel^{-1}s^{-1}$) and readout noise ($5e^{-}$ assuming up-the-ramp readout) for typical H$2$RGs. The SNR as a function of the exposure time is shown in Fig.~\ref{fig:SNR}. 

The plot shows that for stars with H-mag$<8$ (I-mag$<12$), a SNR per pixel of $100/50$ can be obtained in $30/10$~s of integration or less with optimum levels of AO correction. Even at the faint end, it is possible to obtain a SNR per pixel of $25$ in $30$~s on a star with H-mag $=11.08$ (I-mag $=13.45$). If however, the performance of the AO system was degraded for the reasons outlined above, it is clear that this would only limit efficient operation on the faintest of stars (I-mag$>12$). For the instrument described here, all targets are photon noise limited. 

The SNR is only one parameter which is used to determine the ultimate precision of the instrument (m/s) and terms like the spectral content (number of lines and the sharpness of the features), stellar noise, and spectrograph stability need to be considered. Indeed, from the simulations in~\cite{reiners2010}, which take into account a SNR per pixel of $\sim100$ at $R=60000$ in the H-band, it would be possible to achieve a spectral-content-limited radial velocity precision of $4$ and $10$ m/s for M$9$ and M$3$ spectral type stars, respectively. Spectral-content-limited means the highest precision achievable, limited by the number of spectral lines over the bandwidth of observation, and takes into account their sharpness. The instrument proposed here can offer a significantly higher SNR per pixel for bright targets or with co-adding of data for faint targets, which means that even higher precisions are achievable with it based on spectral content alone. 

One could replace the AWG with an Echelle grating and cross-disperser instead. In this case, the throughput of the spectrograph would be similar, yielding a similar SNR on-sky.

Since imperfect scrambling and modal noise are eliminated in this design, these noise terms will not affect the final precision. Further, it is possible to simultaneously inject the light from a calibration source into the AWG via a second SMF, which can be separated on the detector by the cross-disperser. Thus, the calibration spectrum and the stellar spectrum are recorded simultaneously, with the calibration data interleaved between the science data on the detector. In this way, drifts in the spectrograph over time can be carefully monitored and calibrated out. By using the laser-locked Fiber Fabry-Perot etalon discussed above, which offers a stability at the several cm/s level, it should be possible to achieve high precision calibration with a photonic-based spectrograph. The precision of such an instrument on cool stars would be entirely limited by the spectral content and stellar noise and it would certainly be capable of achieving several m/s, the level required for detecting an exo-Earth in the HZ of a nearby M-dwarf. This demonstrates that a SMF-fed photonic-based spectrograph would be efficient enough to be useful for high-precision studies of cool stars. Of course, this instrument could also be used in a post-coronagraphic mode as outlined above and in this case it could offer insights into the chemical composition of the companion itself while helping to further constrain the orbital properties of both the host and companion. 

\section{Summary}\label{sec:summary}
The high Strehl ratios ($\sim90\%$) and unprecedented PSF stability provided by ExAO systems can be used to advance stellar spectroscopy. With these properties, efficient direct injection of light into single-mode fibers or few port photonic lanterns becomes possible, eliminating imperfect input scrambling and modal noise and hence improving PSF calibration and spectrograph precision. Once the light is efficiently coupled to a diffraction-limited device it is possible to exploit a host of photonic functionalities with great efficiency, including spectral filters~\citep{trinh2013} and compact calibration combs~\citep{schwab15}. This enables miniaturized instruments to be developed that could be placed directly in the focal plane of an ELT. These instruments can be extremely compact and robust and they can be mass produced. Other features of ExAO facilities can also enable new capabilities like spatially resolved spectroscopy of close-separation binaries or even post-coronagraphic spectroscopy. By reducing the photon noise, it will be possible to detect and study molecular features in the atmospheres of giant planets in greater detail. ExAO systems are very versatile and will undoubtedly find applications in many fields other than their intended one of high contrast imaging. The key is to identify their potential and harness it.

\acknowledgments
This research was supported by the Australian Research Council Centre of Excellence for Ultrahigh bandwidth Devices for Optical Systems (project number CE110001018). The authors acknowledge support from the JSPS (Grant-in-Aid for Research \#$23340051$ \& \#$26220704$). We would like to thank Yulia Gurevich for help with the manuscript.

\end{document}